\documentclass[english,showpacs,floatfix,12pt]{revtex4}
\usepackage[dvips]{graphicx}
\usepackage{longtable}
\usepackage{amsmath,amssymb}
\usepackage{dcolumn}
\usepackage{latexsym}
\usepackage{babel}
\usepackage[latin1]{inputenc}
\usepackage{color}
\usepackage[colorlinks]{hyperref}
\def\eps{\epsilon}
\def\pa{\partial}
\def\al{\alpha}
\def\ga{\gamma}
\def\Ga{\Gamma}
\def\dl{\delta}
\def\la{\lambda}
\def\La{\Lambda}
\def\be{\beta}
\def\kp{\kappa}
\def\th{\theta}

\def\sg{\sigma}
\def\vf{\varphi}
\def\vr{\varrho}
\def\om{\omega}
\def\Om{\Omega}
\def\Up{\Upsilon}
\def\nb{\nabla}
\def\lan{\langle}
\def\ran{\rangle}
\def\l{\left}
\def\r{\right}
\def\nn{\nonumber}
\def\ck{\check}
\def\wt{\widetilde}
\def\diag{\mbox{diag}}
\def\L{{\cal L}}
\def\E{{\cal E}}

\def\sgn{\mbox{sgn}}
\def\H{\textbf{H}}

\begin{document}
\title{\Large\bf The Vierbein Formalism \\
and Energy-Momentum Tensor of Spinors}
\author{Ying-Qiu Gu}
\email{yqgu@fudan.edu.cn} \affiliation{School of Mathematical
Science, Fudan University, Shanghai 200433, China} \pacs{04.20.Cv,
04.20.Fy, 04.20.-q, 11.10.-z}
\date{24th November 2017}

\begin{abstract}
To study the coupling system of space-time and Fermions, we need the
explicit form of the energy-momentum tensor of spinors. The
energy-momentum tensor is closely related to the tetrad frames which
cannot be uniquely determined by the metric. This flexibility
increases difficulties to derive the exact expression and easily
leads to ambiguous results. In this paper, we give a detailed
derivation for the energy-momentum tensor of Weyl and Dirac spinors.
From the results we find that, besides the usual kinetic energy
momentum term, there are three kinds of other additional terms. One
is the nonlinear self-interactive potential, which acts like
negative pressure. The other reflects the interaction of momentum
$p^\mu$ with tetrad. The third is the spin-gravity coupling term
which is a higher order infinitesimal in weak field, but may be
important in a neutron star.  This term is also closely related with
magnetic field of a celestial body.  These results are based on the
decomposition of usual spin connection into geometrical part and
dynamical part, which not only makes calculation simpler, but also
highlights their different physical meanings. In addition, we get a
new tensor $S^{\mu\nu}_{ab}$ in calculation of tetrad formalism,
which plays an important role in the interaction of spinor with
gravity.

\vskip3mm\noindent {Keywords: {vierbein, tetrad, spinor connection,
spinor structure, energy-momentum tensor}}
\end{abstract}

\maketitle

\section{Introduction}
\setcounter{equation}{0}

The spinor field can give explanation for the accelerating expansion
of the universe, and may be a possible candidate of dark matter. It
is studied by some researchers in recently
years\cite{1,2,3,4,gu3,5}.  In these works, the space-time is
usually Friedmann-Lemaitre-Robertson-Walker type with diagonal
metric. The energy-momentum tensor({\bf EMT}) $T_{\mu\nu}$ of
spinors is simple and can be directly derived from Lagrangian of the
spinor field in this case. To consider the interaction of spinor and
gravity in general case is as early as H. Weyl in 1929\cite{weyl}.
There are some approaches to the general expression of EMT of
spinors in curved space-time\cite{3,Brill,Weldon,Zhang}. But the
formalisms are usually quite complicated for practical calculation
and different from each other. In \cite{Brill,Weldon}, according to
the Pauli's theorem
\begin{eqnarray}
\dl \wt\ga^\al=\frac 1 2 \wt\ga_\be \dl g^{\al\be}+[\wt\ga^\al,M]
,\label{1.17}\end{eqnarray} where $M$ is a traceless matrix related
to the frame transformation, the EMT for Dirac spinor $\phi$ was
derived as follows,
\begin{eqnarray}
T^{\mu\nu}=\frac 1 2 \Re\lan \phi^{\dag~}(\wt \ga^\mu i\nb^\nu+\wt
\ga^\nu i\nb^\mu)\phi\ran,\label{1.18}\end{eqnarray} where
$\phi^{\dag~}=\phi^+\ga$ is the Dirac conjugation, $\nb^\mu$ is the
usual covariant derivatives for spinor. A detailed calculation for
variation of action was performed in \cite{3}, and the results were
a little different from (\ref{1.17}) and (\ref{1.18}).

The following calculation shows that, $M$ is still related with $\dl
g^{\mu\nu}$, and provides nonzero contribution to $T^{\mu\nu}$ in
general cases. Besides, the covariant derivatives operator
$i\nb_\mu$ is not parallel to the classical momentum $p_\mu$ and the
expression (\ref{1.18}) is complicated for practical calculation and
some important effects are concealed by the compact form. The exact
EMT of spinor was actually not obtained before.

The derivation of $T_{\mu\nu}$ is quite difficult due to
non-uniqueness representation and complicated formalism of vierbein
or tetrad frames. In this paper, we give a systematical and detailed
calculation for EMT of Weyl and Dirac spinors. We clearly establish
the relations between tetrad and metric at first, and then solve the
Euler derivatives with respect to $g_{\mu\nu}$ to get explicit and
rigorous $T_{\mu\nu}$.

From the results we find some new and interesting conclusions.
Besides the usual kinetic energy momentum term, we find three  kinds
of other additional terms in EMT of bispinor. One is the self
interactive potential, which acts like negative pressure. The other
reflects the interaction of momentum $p^\mu$ with tetrad, which
vanishes in classical approximation. The third is the spin-gravity
coupling term $\Om_\al s^\al$, which is a higher order infinitesimal
in weak field, but becomes important in a neutron star.  This term
is the eye of a particle with location and navigation functions, and
is closely related with magnetic field of a celestial body.  All
these results are based on the decomposition of usual spin
connection $\Ga_\mu$ into geometrical part $\Up_\mu$ and dynamical
part $\Om_\mu$, which not only makes calculation simpler, but also
highlights their different physical meanings. In addition, in the
calculation of tetrad formalism we find a new tensor
$S^{\mu\nu}_{ab}$ which plays an important role in the interaction
of spinor with gravity and appears in many places.

The materials are organized as follows: In the next section, we
specify all notations, conventions and relevant equations used in
the discussion. In the third section, we provide the technical
foundations for the following derivation of EMT. We derive the exact
EMT of spinor and its classical approximation in section IV, and
then we give some simple discussions and illustrations in the last
section.

\section{notations and equations}
\setcounter{equation}{0}

Denote the metric  and coordinates of the Minkowski space-time
respectively by
\begin{eqnarray}
\eta_{ab} = \diag(-1,-1,-1,~1),\quad \dl X^\mu=(\dl X,\dl Y,\dl
Z,c\dl T).
\end{eqnarray} The Pauli matrices are expressed by
\begin{eqnarray}
 \sg^{\mu}&=& \left \{\left(\begin{array}{cc}
 0 & 1 \\  1 & 0\end{array}\right),\left(\begin{array}{cc}
 0 & -i \\   i & 0\end{array}\right),\left(\begin{array}{cc}
 1 & 0 \\   0 & -1\end{array}\right),\left(\begin{array}{cc}
 1 & 0 \\   0 & 1\end{array}\right)
 \right\},\label{1.1}\\
\wt{\sg}^{\mu}&=&(-\vec\sg,~\sg^0),\qquad
\vec\sg=(\sg^1,\sg^2,\sg^3). \label{1.2}
\end{eqnarray} where ${a,\mu}\in{0,1,2,3}$.  In this paper, we use the
Greek characters stand for indices of curvilinear coordinates and
Latin characters for indices of local Minkowski coordinates. The
element of the space-time can be expressed by
\begin{eqnarray} d {\bf x}=\wt\ga_\mu dx^\mu=\ga_a \dl X^a,
\label{qnt}\end{eqnarray} where  $\ga_a$ and $\wt \ga^\mu$ are
tetrad expressed by Dirac matrices
\begin{eqnarray}
 \ga^{\mu}= \left(\begin{array}{cc}
 0 & \wt \sg^\mu \\  \sg^\mu & 0\end{array}\right), \quad \wt \ga^\mu=h^\mu_{~a}\ga^a,\quad \wt \ga_\mu=l_\mu^{~a}\ga_a,\label{gam}
\end{eqnarray}
which satisfies the  $C\ell(1,3)$ Clifford
algebra\cite{sachs,Newman,Bade,Bergm,Crawf}
\begin{eqnarray}\ga_\mu\ga_\nu+\ga_\nu\ga_\mu=\eta_{\mu\nu},\qquad \wt\ga_\mu\wt\ga_\nu+\wt\ga_\nu\wt\ga_\mu=g_{\mu\nu}. \label{1.10*}\end{eqnarray}

In Minkowski space-time, we have Dirac equation
\begin{eqnarray}
\ga^\mu i\pa_\mu\phi=m \phi, \label{1.4}
 \end{eqnarray}
or in chiral form\cite{sachs}
\begin{eqnarray}\left\{\begin{array}{l}
\sg^\mu i\pa_\mu\psi=m\wt\psi,\\
\wt\sg^\mu i\pa_\mu\wt\psi=m\psi,
\end{array}\right. \qquad  \phi= \left(\begin{array}{c}
 \psi \\ \wt\psi\end{array}\right), \label{1.6}\end{eqnarray}
where $\psi,~\wt\psi$ are two Weyl spinors.

Denote the Pauli matrices in curved space-time by
\begin{eqnarray}\left\{\begin{array}{l}
\vr^\mu=h^\mu_{~a}\sg^a,\quad\vr_\mu=l_\mu^{~a}\sg_a, \\
\wt\vr^\mu=h^\mu_{~a}\wt\sg^a, \quad \wt\vr_\mu=l_\mu^{~a}\wt\sg_a,
\end{array}\right.\qquad \wt\ga^\mu = \left(\begin{array}{cc} 0 & \wt \vr^\mu \\ \vr^\mu & 0
\end{array}\right),\label{1.9}\end{eqnarray}
then we have Clifford algebra as follows
\begin{eqnarray}\vr^\mu\wt\vr^\nu+\vr^\nu\wt\vr^\mu=\wt\vr^\mu\vr^\nu+\wt\vr^\nu\vr^\mu=2
g^{\mu\nu}. \label{1.10}\end{eqnarray} The Weyl spinor equation
(\ref{1.6}) in curved space-time becomes
\begin{eqnarray}\left\{\begin{array}{l}
\vr^\mu i(\pa_\mu+\Ga_\mu)\psi=m\wt\psi,\\
\wt\vr^\mu i(\pa_\mu+\wt\Ga_\mu)\wt\psi=m\psi,
\end{array}\right. \label{1.11}\end{eqnarray}
where $\Ga_\mu$ and $\wt\Ga_\mu$ are the spinor affine
connections\cite{1,2,3,4,5,Bergm,Weldon},
\begin{eqnarray}
\Ga_\mu=\frac 1 4 \wt\vr_\nu\vr^\nu_{;\mu},\qquad \wt\Ga_\mu=\frac 1
4 \vr_\nu\wt\vr^\nu_{;\mu}, \label{gg1.11}\end{eqnarray} in which
$\vr^\mu_{;\nu}=\pa_\nu\vr^\mu+\Ga^\mu_{\al\nu}\vr^\al$.

In order to disclose the physical meanings of connection, the total
covariant derivatives of spinor can be represented in the following
form\cite{gu2},
\begin{eqnarray}
\vr^\mu i(\pa_\mu +\Ga_\mu)=\vr^\mu\l[i(\pa_\mu+\Up_\mu)+\Om_\mu\r],\label{(omg)}\\
\wt \vr^\mu i(\pa_\mu +\wt
\Ga_\mu)=\wt\vr^\mu\l[i(\pa_\mu+\Up_\mu)-\Om_\mu\r].\label{(omg1)}\end{eqnarray}
In which, $\Up_\mu$ is related with the grade-1 Clifford algebra,
which has only geometrical effect,
\begin{eqnarray}
\Upsilon_\mu\equiv \frac 1 2 (l_\mu^a\pa_\nu
h^\nu_a+\pa_\mu\ln\sqrt{g})= \frac 1 2 h^\nu_a(\pa_\mu
l_\nu^a-\pa_\nu l_\mu^a). \label{ops}\end{eqnarray} But $\Om_\mu$ is
related with grade-3 Clifford algebra\cite{nst}. Similarly, for
Dirac bispinor in curved space-time we have
\begin{eqnarray}
\nb_\mu\phi=(\pa_\mu+\hat\Ga_\mu)\phi,\qquad \hat\Ga_\mu=\frac 1 4
\wt\ga_\nu\wt\ga^\nu_{;\mu}. \label{1.16}\end{eqnarray}

More generally, the Lagrangian corresponding to (\ref{1.11}) with
electromagnetic interaction and nonlinear self-potential $F$ is
given by
\begin{eqnarray}
 {\cal L}_m &=& \Re  \lan \phi^+\wt\al^\mu \hat p_\mu\phi \ran +\Om_\mu \phi^+
\hat s^\mu \phi -m \phi^+\ga_0\phi+F,\label{ldrc}\\
&=& \Re  \lan \psi^+\vr^\mu \hat p_\mu\psi+\wt\psi^+\wt\vr^\mu \hat
p_\mu\wt\psi \ran +\psi^+\Om \psi+\wt\psi^+\wt\Om
\wt\psi-m(\wt\psi^+\psi+\psi^+\wt\psi)+F, \label{3.2}
\end{eqnarray} where $\Re\lan\ran$ means taking real part, $\wt w>0$ is the nonlinear coupling coefficient,
\begin{eqnarray}
F=\frac 1 2 \wt w\ck\ga^2,\qquad \ck\ga= \phi^+\ga_0\phi.
\label{ptn}
\end{eqnarray}
$(\hat
p_\mu, \wt\al^\mu, \hat s^\mu)$ are respectively momentum, current
and spin operators defined by
\begin{eqnarray}
\hat p_\mu= i(\pa_\mu+\Up_\mu)-e A_\mu,\quad
\wt\al^\mu=\diag(\vr^\mu,\wt\vr^\mu),\quad \hat
s^\mu=\diag(\vr^\mu,-\wt\vr^\mu).\label{mtnp}
\end{eqnarray}
In local central coordinate system $\dl X^a$, $\hat s^a$ equals to
$\frac 1 2 \hbar$ in one direction but vanishes in other direction.
$\Om$ and $\wt\Om$ are two Hermitian matrix defined by
\begin{eqnarray}\left\{\begin{array}{l}
\Om\equiv  \frac i 8[\vr^\mu
\wt\vr^\al\pa_\mu\vr_\al-(\pa_\mu\vr_\al)\wt\vr^\al\vr^\mu]=\Om_\mu\vr^\mu=\om_a\sg^a,\\
\wt\Om\equiv  \frac i 8[\wt\vr^\mu
\vr^\al\pa_\mu\wt\vr_\al-(\pa_\mu\wt\vr_\al)\vr^\al\wt\vr^\mu]=-\Om_\mu\wt\vr^\mu=-\om_a\wt\sg^a.
\end{array}\right.\label{omga}
\end{eqnarray}
For any diagonal metric, it easy to check $\Om=\wt\Om=0$. By
straightforward calculation we have\cite{gu2}
\begin{eqnarray}\left\{\begin{array}{l}
\om_0 = -\frac 1 4 (\vec h^\al\times\vec h^\be)\cdot\pa_\al \vec
l_\be,\\
\vec\om =-\frac 1 4 \l(\pa_\al l_\be^{~0}(\vec h^\al\times\vec
h^\be)-(h^\al_{~0}\vec h^\be-h^\be_{~0}\vec
h^\al)\times\pa_\al\vec l_\be \right),\\
\Om_\mu = -\frac 1 4 \left((\vec h^\al\times\vec
h^\be)\cdot(l_\mu^{~0}\pa_\al \vec l_\be-\vec l_\mu\pa_\al
l_\be^{~0})+\vec l_\mu\cdot[(h^\al_{~0}\vec h^\be-h^\be_{~0}\vec
h^\al)\times\pa_\al\vec l_\be]\right),
\end{array}\right.
\label{gg2.11}
\end{eqnarray} in which $\vec \om=(\om^1,\om^2,\om^3)$. (\ref{gg2.11}) defines the dynamical part of the spinor connection.

\section{relations between Tetrad and Metric}
\setcounter{equation}{0} In this section, we give an explicit
representation of tetrad formalism. The derivation of EMT is based
on this representation. Different from the cases of vector and
tensor, in general relativity the dynamical equations for spinor
fields depend on the local tetrad frame, which make the
representation of the spinor connection and the EMT quite
complicated. Assume that $x^\mu=(x,y,z,ct)$ is the coordinates  and
$\dl X^a$ is the element vector in the tangent space at fixed point
$x^\mu$. The tetrad $\wt\ga^\mu$ cannot be uniquely determined by
metric, which can be only determined to an arbitrary local Lorentz
transformation. Now we derive some important relations used below.

{\bf Lemma 1.} {\em For Pauli matrices (\ref{1.1}) and (\ref{1.2}),
we have relations
\begin{eqnarray}\left\{\begin{array}{l}
 \sg^a\wt\sg^b\sg^c-\sg^c\wt\sg^b\sg^a= 2i\eps^{abcd}\sg_d,\\
 \wt\sg^a\sg^b\wt\sg^c-\wt\sg^c\sg^b\wt\sg^a=
 -2i\eps^{abcd}\wt\sg_d,
 \end{array}\right.
\label{clf}
\end{eqnarray}
in which $\eps^{abcd}$ is the permutation function.}

Lemma 1 can be easy checked. We have only 4 nonzero cases for each
equation.

For metric $g_{\mu\nu}$, not losing generality we assume that, in
the neighborhood of $x^\mu$, $dx^0$ is time-like and $(dx^1, dx^2,
dx^3)$ are space-like. This means $g_{00}\ge 0$ and $g_{kk}\le
0(k\ne 0 )$, and the following definitions of $J_k$ are real numbers
\begin{eqnarray} J_1=\sqrt {-g_{{1 1}}},~ J_2=\sqrt { \left| \begin
{array}{cc} g_{{1 1}}&g_{{1 2}}
\\  g_{{2 1}}&g_{{2 2}}\end {array} \right| },~
J_3=\sqrt {- \left| \begin {array}{ccc} g_{{1 1}}&g_{{1 2}}&g_{{1
3}}
\\  g_{{2 1}}&g_{{2 2}}&g_{{2 3}}\\  g
_{{3 1}}&g_{{3 2}}&g_{{3 3}}\end {array} \right| }
,~J_0=\sqrt{-\det(g)}. \label{2.2}\end{eqnarray} Denote
\begin{eqnarray}
u_1=\left| \begin {array}{cc} g_{{1 1}}&g_{{1 2}}\\
 g_{{3 1}}&g_{{3 2}}\end {array} \right|, \quad
u_2=\left| \begin {array}{cc} g_{{1 1}}&g_{{1 2}}\\
 g_{{0 1}}&g_{{0 2}}\end {array} \right|,\quad
u_3= \left| \begin {array}{cc} g_{{2 1}}&g_{{2 2}}\\
 g_{{ 3 1}}&g_{{3 2}}\end {array} \right|,
\label{2.3}\end{eqnarray} and
\begin{eqnarray}
v_1=\left| \begin {array}{ccc} g_{{1 2}}&g_{{1 3}}&g_{{1 0}}\\
 g_{{2 2}}&g_{{2 3}}&g_{{2
0}}\\ g _{{3 2}}&g_{{3 3}}&g_{{3 0}}\end {array} \right|,~~v_2=
\left|
\begin {array}{ccc} g_{{1 1}}&g_{{1 3}}&g_{{1 0}}
\\  g_{{2 1}}&g_{{2 3}}&g_{{2 0}} \\  g
_{{3 1}}&g_{{3 3}}&g_{{3 0}}\end {array} \right|,~~
v_3=\left|\begin {array}{ccc} g_{{1 1}}&g_{{1 2}}&g_{{1 0}}\\
 g_{{2 1}}&g_{{2 2}}&g_{{2 0}}\\
 g _{{3 1}}&g_{{3 2}}&g_{{3 0}}\end {array}
\right|, \label{2.4}\end{eqnarray} then we have the following
conclusion.

 {\bf Theorem 2.} {\em For LU decomposition of matrix
$(g_{\mu\nu})$,
\begin{eqnarray}
(g_{\mu\nu})=L(\eta_{ab}) L^+,~~(g^{\mu\nu})=U(\eta_{ab})
U^+,~~U=L^*=(L^+)^{-1}, \label{2.5}\end{eqnarray} with positive
diagonal elements, we have the following unique solution}
\begin{eqnarray}
L=(L_\mu^{~a})=\left( \begin {array}{cccc} -{\frac {g_{{1 1}}}{{J_1}}}&0&0&0\\
 -{\frac {g_{{2 1}}}{{
J_1}}}&{\frac {{ J_2}}{{  J_1}}}&0&0\\
 -{\frac {g_{{3 1}}}{{ J_1}}}&{ \frac {{ u_1}}{{
J_1}\, { J_2}}}&{\frac {{ J_3}}{{ J_2}}}&0
\\  -{\frac {g_{{0 1}}}{{ J_1}}}&{\frac {{ u_2}}{{
 J_1}\,{ J_2}}}&-{\frac {{ v_3}}{{ J_2}\,{
J_3}}}&{\frac {{  J_0}}{{ J_3}}}\end {array} \right),
\label{2.6}\end{eqnarray}
\begin{eqnarray}
U=(U^\mu_{~a})=\left( \begin {array}{cccc} {{\frac 1 {J_1}}} &{\frac
{g_{{2 1}}}{{ J_1}\,{ J_2}}}&{\frac {{ u_3}}{{ J_2}\,{
J_3}}}&{\frac {{ v_1 }}{{ J_3}\,{ J_0}}}\\
 0&{\frac {{ J_1}}{{ J_2 }}}&-{\frac {{ u_1}}{{
J_2}\,{ J_3}}}&-{\frac {{ v_2}}{{ J_3 }\,{ J_0}}}\\
 0&0&{\frac
{{ J_2}}{{ J_3}}}&{ \frac {{ v_3}}{{ J_3}\,{ J_0}}}\\
 0&0&0&{\frac {{ J_3}}{{ J_0}}}\end {array}
\right). \label{2.7}\end{eqnarray} (\ref{2.6}) and (\ref{2.7}) can
be checked directly. For any other solutions of (\ref{gam}), we have

{\bf Theorem 3.} {\em For any solution of tetrad (\ref{gam}) in
matrix form $(l_\mu^{~a})$ and $(h^\mu_{~a})$, there exists a local
Lorentz transformation $\dl {X'}^a=\La^a_{~b}\dl X^b$ independent of
$g_{\mu\nu}$, such that
\begin{eqnarray}
(l_\mu^{~a})=L \La^+,\qquad (h^\mu_{~a})=U \La^{-1},
\label{2.8}\end{eqnarray} where $\La=(\La^a_{~b})$ stands for the
matrix of Lorentz transformation.}

Theorem 3 can be checked as follows,
\begin{eqnarray}
(g_{\mu\nu})=L(\eta_{ab}) L^+=(l_\mu^{~a})(\eta_{ab})
(l_\mu^{~a})^+~\Longrightarrow~L^{-1}(l_\mu^{~a})(\eta_{ab})
(L^{-1}(l_\mu^{~a}))^+=(\eta_{ab}). \label{2.9}\end{eqnarray} Then
we have a Lorentz transformation matrix $\La=(\La^a_{~b})$, such
that
\begin{eqnarray}
L^{-1}(l_\mu^{~a})=\La^+~\Longrightarrow~(l_\mu^{~a})=L
\La^+,~~{\mbox{or}}~~{l}_\mu^{~a}=L_\mu^{~b} \La^a_{~b}.
\label{2.10}\end{eqnarray} Similarly we have $(h^\mu_{~a})=U
\La^{-1}$.

{\bf Remark.} {\em The decomposition (\ref{2.5}) is nothing but a
Gram-Schmidt orthogonalization for vectors $dx^\mu=(dx,dy,dz,dt)$ in
tangent space-time in the order $dt\to dz \to dy \to dx$, namely in
vector form $\dl X = dx L$ or}
\begin{eqnarray}
ds^2&=& g_{\mu\nu} dx^\mu dx^{\nu} = \eta_{ab} \dl X^a \dl X^b \nn\\
&=& -(L_x^{~X}dx+L_y^{~X}dy+L_z^{~X}dz+L_t^{~X}dt)^2\nn\\
&~&-(L_y^{~Y}dy+L_z^{~Y}dz+L_t^{~Y}dt)^2-(L_z^{~Z}dz+L_t^{~Z}dt)^2+(L_t^{~T}dt)^2.
\label{shcmidt}\end{eqnarray} (\ref{shcmidt}) is a direct result of
(\ref{2.6}), but (\ref{shcmidt}) manifestly shows the geometrical
meanings of the tetrad components $L_\mu^{~a}$. Obviously,
(\ref{shcmidt}) is much convenient for practical calculation of
tetrad parameters.

Define a spinor coefficient tensor by
\begin{eqnarray}
S^{\mu\nu}_{ab} \equiv \frac 1 2
(h^\mu_{~a}h^\nu_{~b}+h^\nu_{~a}h^\mu_{~b})\sgn(a-b).
\label{2.13*}\end{eqnarray} $S^{\mu\nu}_{ab}$ is symmetrical for
Riemann indices $(\mu,\nu)$ but anti-symmetrical for Minkowski
indices $(a,b)$. (\ref{2.13*}) is important for the following
calculation. By representation of (\ref{2.6}), (\ref{2.7}) and
relation (\ref{2.8}), we can check the following results by
straightforward calculation.

{\bf Theorem 4.} {\em For any solution  of tetrad (\ref{gam}), we
have
\begin{eqnarray}
\frac {\pa l_\al^{~n}}{\pa g_{\mu\nu}}&=&\frac 1 4 (\dl^\mu_\al
h^\nu_{~m}+\dl^\nu_\al h^\mu_{~m})\eta^{nm}+\frac 1 2
S^{\mu\nu}_{ab}l_\al^{~a}\eta^{nb}. \label{2.14}\\
\frac {\pa h^\al_{~a}}{\pa g_{\mu\nu}} &=&-\frac 1 4 (h^\mu_{~a}
g^{\al\nu}+h^\nu_{~a} g^{\mu\al})-\frac 1 2
S^{\mu\nu}_{ab}h^\al_{~n}\eta^{nb}. \label{2.14*}
\end{eqnarray}
Or equivalently,
\begin{eqnarray}
\frac {\pa \vr_\al}{\pa g_{\mu\nu}} &=&\frac 1 4
(\dl^\mu_\al\vr^\nu+\dl^\nu_\al\vr^\mu)+\frac 1 2
S^{\mu\nu}_{ab}l_\al^{~a}\sg^b. \label{2.13} \\
\frac {\pa \vr^\al}{\pa g_{\mu\nu}} &=&-\frac 1 4
(g^{\mu\al}\vr^\nu+g^{\nu\al}\vr^\mu)-\frac 1 2
S^{\mu\nu}_{ab}h^\al_{~n}\eta^{nb}\sg^a. \label{2.15*}
\end{eqnarray}
In (\ref{2.14})-(\ref{2.15*}) we set $\frac {\pa \vr_\al}{\pa
g_{\mu\nu}}=\frac {\pa \vr_\al}{\pa g_{\nu\mu}}=\frac 1 2 \frac {d
\vr_\al}{d g_{\mu\nu}}~(\mu\neq\nu)$ to get the tensor form. $\frac
{d ~~}{d g_{\mu\nu}}$ means the total derivative for $g_{\mu\nu}$
and $g_{\nu\mu}$. For given vector $A^\al$, we have}
\begin{eqnarray}
A^\al \frac {\pa \vr_\al}{\pa g_{\mu\nu}} &=&\frac 1 4 (A^\mu
\vr^\nu+A^\nu \vr^\mu)+\frac 1 2
S^{\mu\nu}_{ab}A^\al l_\al^{~a}\sg^b. \label{ar1} \\
A_\al \frac {\pa \vr^\al}{\pa g_{\mu\nu}} &=&-\frac 1 4
(A^{\mu}\vr^\nu+A^{\nu} \vr^\mu)-\frac 1 2 S^{\mu\nu}_{ab}A_\al
h^\al_{~n}\eta^{nb}\sg^a. \label{ar2}
\end{eqnarray}

 Similarly, for Dirac matrices we have

{\bf Corollary 5.} {\em  For Dirac  matrices (\ref{gam}), we have
\begin{eqnarray}
\frac {\pa {\wt\ga}_\al}{\pa g_{\mu\nu}}=\frac 1 4 (\dl^\mu_\al
{\wt\ga}^\nu+\dl^\nu_\al {\wt\ga}^\mu)+\frac 1 2 S^{\mu\nu}_{ab}
l_\al^{~a}\ga^b.
\end{eqnarray} Or equivalently, for  any given vector $A^\mu$, we
have}
\begin{eqnarray}
A^\al\frac {\pa {\wt\ga}_\al}{\pa g_{\mu\nu}}=\frac 1 4
(A^\mu{\wt\ga}^\nu+A^\nu{\wt\ga}^\mu)+\frac 1 2 S^{\mu\nu}_{ab}A^\al
l_\al^{~a}\ga^b.
\end{eqnarray}

\section{Energy-momentum tensor of spinors}
\setcounter{equation}{0}

Now we consider the coupling system of spinor and gravity. The Ricci
tensor and scalar curvature is defined by
\begin{eqnarray}
R_{\mu\nu}=\pa_\mu\Ga^\al_{\nu\al}-\pa_\al\Ga^\al_{\mu\nu}+\Ga^\al_{\mu\be}\Ga^\be_{\nu\al}-
\Ga^\al_{\mu\nu}\Ga^\be_{\al\be},\qquad R\equiv
g^{\mu\nu}R_{\mu\nu}. \nonumber
\end{eqnarray}
The total Lagrangian of the system reads
\begin{eqnarray}
{\cal L}=\frac 1 {2\kp}{\cal L}_g+{\cal L}_m=-\frac1
{2\kp}(R-2\Lambda)+{\cal L}_m, \label{lag}
\end{eqnarray}
where $\kp=8\pi G$, $\Lambda$ is the cosmological constant, ${\cal
L}_m$ the Lagrangian of spinors (\ref{3.2}). Variation of the
Lagrangian (\ref{lag}) with respect to $g_{\mu\nu}$ we get
Einstein's equation
\begin{eqnarray}
R^{\mu\nu}-\frac 1 2g^{\mu\nu}R+\La g^{\mu\nu}=\kp T^{\mu\nu},
\end{eqnarray}
where $T^{\mu\nu}$ is EMT of the spinors, which is defined by
\begin{eqnarray}
T^{\mu\nu}=-2\frac {\dl({\cal L}_m \sqrt{g})}{\sqrt{g}\dl
g_{\mu\nu}}=-2\l(\frac {\pa{\cal L}_m }{\pa g_{\mu\nu}}-(\pa_\al
+\Ga^\ga_{\al\ga})\frac {\pa{\cal L}_m }{\pa(\pa_\al
g_{\mu\nu})}\r)-g^{\mu\nu}{\cal L}_m ,\label{3.3}
\end{eqnarray}
where $g=|\det(g_{\mu\nu})|$ and $\frac {\dl~~} {\dl g_{\mu\nu}}$ is
the Euler derivatives.

We take $\psi$ as example to derive relations, because we have
similar results for $\wt\psi$. By  (\ref{3.2}) and (\ref{2.15*}) we
have
\begin{eqnarray}
\E^{\mu\nu}_p\equiv \frac \pa {\pa g_{\mu\nu}}  \Re \lan\psi^+
\vr^\al \hat p_\al \psi \ran = -\frac 1 4\Re \lan\psi^+(\vr^\mu \hat
p^\nu +\vr^\nu \hat p^\mu + 2 S^{\mu\nu}_{a
b}h^\al_{~n}\eta^{nb}\sg^a\hat p_\al)\psi \ran . \label{eng0}
\end{eqnarray}
By (\ref{omga}), (\ref{clf}) and (\ref{ar1}) we rewrite $\Om$ as
follows,
\begin{eqnarray}
\Om &=& \frac i 8 \l(\vr^\mu\wt\vr^\al\frac {\pa\vr_\al}{\pa
g_{\la\kp}}-\frac {\pa\vr_\al}{\pa
g_{\la\kp}}\wt\vr^\al\vr^\mu\r)\pa_\mu g_{\la\kp},\nn\\
&=& \frac i {16} \l(\sg^a\wt\sg^b\sg^c-\sg^c\wt\sg^b\sg^a\r)h^\mu_{~a}S^{\la\kp}_{bc}\pa_\mu g_{\la\kp},\qquad\qquad\qquad~~ {\rm by~~(\ref{ar1})} \nn\\
&=& -\frac 1 8 \eps^{abcd}\sg_d h^\mu_{~a}S^{\la\kp}_{bc}\pa_\mu
g_{\la\kp}=\frac 1 8 \eps^{dabc}\sg_d
h^\mu_{~a}S^{\la\kp}_{bc}\pa_\mu g_{\la\kp}.\qquad {\rm
by~~(\ref{clf})}
 \label{omga1}
\end{eqnarray}
By (\ref{omga1}), we get $(\Om^\al,\om^a)$ expressed by $\pa_\al
g_{\mu\nu}$ as follows,
\begin{eqnarray}
\om^d = \frac 1 8 \eps^{dabc} h^\al_{~a}S^{\mu\nu}_{bc}\pa_\al
g_{\mu\nu},\qquad \Om^\al = \frac 1 8 \eps^{dabc}
h_{~d}^{\al}h^\be_{~a}S^{\mu\nu}_{bc}\pa_\be g_{\mu\nu}.
 \label{omOm}
\end{eqnarray}
Then we get
\begin{eqnarray}
\frac {\pa(\psi^+\Om\psi)} {\pa g_{\mu\nu}} &=& -\frac 1 8
\eps^{abcd} \ck\sg_d \frac {\pa(h^\al_{~a}S^{\la\kp}_{bc})}{\pa
g_{\mu\nu}} \pa_\al g_{\la\kp}  ,\qquad \ck\sg_d\equiv \psi^+\sg_d
\psi,
\label{dt0} \\
(\pa_\al+ \Ga^\be_{\al\be}) \frac {\pa(\psi^+\Om\psi)} {\pa(\pa_\al
g_{\mu\nu})} &=& -\frac 1 8 \eps^{abcd} \l( h^\al_{~a}
S^{\mu\nu}_{bc} (\pa_\al+\Ga^\be_{\al\be}) \ck\sg_d +\ck\sg_d \frac
{\pa(h^\al_{~a} S^{\mu\nu}_{bc})}{\pa g_{\la\kp}} \pa_\al g_{\la\kp}
\r), \label{dt1}
\end{eqnarray}
and then we have the Euler derivative for $\psi^+\Om\psi$ as
\begin{eqnarray}
\E^{\mu\nu}_\Om &\equiv &\frac {\dl(\psi^+\Om\psi)}{\dl
g_{\mu\nu}}=\frac {\pa(\psi^+\Om\psi)} {\pa g_{\mu\nu}} -(\pa_\al+
\Ga^\be_{\al\be})
\frac {\pa(\psi^+\Om\psi)} {\pa(\pa_\al g_{\mu\nu})}, \nn\\
&=& \frac 1 8 \eps^{abcd} \l[ h^\al_{~a} S^{\mu\nu}_{bc}
(\pa_\al+\Ga^\be_{\al\be}) \ck\sg_d +\ck\sg_d\l( \frac
{\pa(h^\al_{~a} S^{\mu\nu}_{bc})}{\pa g_{\la\kp}}- \frac
{\pa(h^\al_{~a} S^{\la\kp}_{bc})}{\pa g_{\mu\nu}} \r) \pa_\al
g_{\la\kp} \r]. \label{tom1}
\end{eqnarray}

In principle, we can substitute (\ref{2.14*}) into (\ref{tom1}) and
then simplify it to get the final expression of $\E^{\mu\nu}_\Om$.
However, the structure of $\E^{\mu\nu}_\Om$ has been clear, and we
need not to do so complicated calculations now, because it is a
second order tenser including only linear first order derivatives
$\pa_\mu\ck\sg_d$ and $\pa_\mu g_{\al\be}$. The vectors and tensors
constructed by metric and its linear first order derivatives are
only $(\Up_\mu, \Om_\mu)$ multiplied by $g_{\mu\nu}$ and
$g^{\al\be}$. Substituting $\ck \sg_d=h^\be_{~d} \ck\vr_\be$ into
(\ref{tom1}), then the simplified expression should take the
following covariant form
\begin{eqnarray}
\E^{\mu\nu}_\Om  = \frac 1 8 \eps^{abcd} h^\al_{~a} h^\be_{~b}
S^{\mu\nu}_{cd} \ck\vr_{\be;\al} + g^{\mu\nu}(k_1\Om^\al+ k_2
\Up^\al)\ck\vr_\al, \label{tom2*}
\end{eqnarray}
where $(k_1,k_2)$ are 2 constants to be determined.

For diagonal metric we have $\Om=S^{\mu\mu}_{ab}=
\E^{\mu\mu}_\Om=0$, and then by (\ref{tom2*}) we get $k_2=0$. In
Gaussian normal coordinate system $g_{\mu\nu}=(-\bar g_{kl},1)$, we
have Hamiltonian for linear bispinor\cite{gu2}
\begin{eqnarray}
\H = -\wt\al^k \hat p_k+eA_0+m \ga_0-\Om_\mu \hat s^\mu.\label{haml}
\end{eqnarray}
By (\ref{3.3}) and (\ref{tom2*}), we have
\begin{eqnarray}
T^{00}=-2(\E^{00}_\Om +\E^{00}_{\wt\Om})+\cdots=-2k_1\Om_\mu \ck
s^\mu+\cdots,\label{ham2}
\end{eqnarray}
in which the omitted terms are  irrelevant with the following
comparison. Since $T^{00}$ is energy of the spinors, in contrast
(\ref{ham2}) with (\ref{haml}) we get $k_1=\frac 1 2$. So we get
\begin{eqnarray}
\E^{\mu\nu}_\Om  = \frac 1 8 \eps^{abcd} h^\al_{~a} h^\be_{~b}
S^{\mu\nu}_{cd} \ck\vr_{\be;\al} +\frac 1 2 g^{\mu\nu}\Om^\al
\ck\vr_\al. \label{tom2}
\end{eqnarray}

Similarly, for $\wt\psi$ we have
\begin{eqnarray}
\E^{\mu\nu}_{\wt \Om}  = -\frac 1 8 \eps^{abcd} h^\al_{~a}
h^\be_{~b} S^{\mu\nu}_{cd} \ck{\wt\vr}_{\be;\al} -\frac 1 2
g^{\mu\nu}\Om^\al \ck{\wt\vr}_\al,\qquad \ck{\wt\vr}_\al\equiv
\wt\psi^+\wt\vr_\al \wt\psi. \label{ep2}
\end{eqnarray}
For total spin $\hat s^\mu$ we have
\begin{eqnarray}
\E^{\mu\nu}_{s}=\E^{\mu\nu}_{\Om}+\E^{\mu\nu}_{\wt \Om}  = \frac 1 8
\eps^{abcd} h^\al_{~a} h^\be_{~b} S^{\mu\nu}_{cd} \ck s_{\be;\al} +
\frac 1 2 g^{\mu\nu}\Om^\al \ck s_\al,\quad \ck s_\al\equiv
\phi^+\hat s\phi. \label{emgs}
\end{eqnarray}
Substituting (\ref{eng0}) and (\ref{emgs}) into  (\ref{3.3}), we get
the EMT for Weyl spinors
\begin{eqnarray}
T^{\mu\nu} &=& \frac 1 2\Re \lan\psi^+(\vr^\mu \hat p^\nu +\vr^\nu
\hat p^\mu)\psi +\wt\psi^+(\wt\vr^\mu \hat p^\nu +\wt \vr^\nu \hat
p^\mu) \wt \psi\ran - g^{\mu\nu}{\cal L}_m\nn\\
&~&   + \Re \lan \psi^+ S^{\mu\nu}_{a b}h^\al_{~n}\eta^{nb}\sg^a\hat
p_\al\psi +\wt\psi^+ S^{\mu\nu}_{a b}h^\al_{~n}\eta^{nb}\wt\sg^a\hat
p_\al\wt\psi \ran \label{tmn}\\
&~&- \frac 1 4 \eps^{abcd} h^\al_{~a} h^\be_{~b} S^{\mu\nu}_{cd} \ck
s_{\be;\al} - g^{\mu\nu}\Om^\al\ck s_\al. \nn\end{eqnarray} For
nonlinear spinor with potential (\ref{ptn}), by Dirac equation we
have $\L_m=-F$.

Substituting the results into (\ref{tmn}), we get the final EMT of
bispinor $\phi$ as follows,
\begin{eqnarray}
T^{\mu\nu} &=& \frac 1 2\Re \lan\phi^+(\wt\al^\mu \hat p^\nu
+\wt\al^\nu \hat p^\mu + 2 S^{\mu\nu}_{a
b}h^\al_{~n}\eta^{nb}\al^a\hat p_\al)\phi \ran  +
g^{\mu\nu}F\nn\\
&~&- \frac 1 4 \eps^{abcd} h^\al_{~a} h^\be_{~b} S^{\mu\nu}_{cd} \ck
s_{\be;\al} - g^{\mu\nu}\Om_\al \ck s^\al. \label{drc}
\end{eqnarray}
The physical meaning of
\begin{eqnarray}
\E^{\mu\nu} \equiv \frac 1 4 \eps^{abcd} h^\al_{~a} h^\be_{~b}
S^{\mu\nu}_{cd} \ck s_{\be;\al}=\frac 1 8 \eps^{abcd} h^\al_{~a}
h^\be_{~b} S^{\mu\nu}_{cd} (\pa_\al \ck s_{\be} - \pa_\be \ck
s_{\al})
\end{eqnarray} is unclear.
Its diagonal components vanish, so it should be very tiny term to
keep $T^{\mu\nu}_{;\nu}=0$.

Now we consider the classical approximation for (\ref{drc}),
\begin{eqnarray}
\phi^+  \wt\al^\nu \phi\to u^\nu\sqrt{1- v^2}\dl^3(\vec x-\vec
X),\quad \hat p^\mu \phi\to m u^\mu \phi,\quad F \to w \sqrt{1-
v^2}\dl^3(\vec x-\vec X).\label{cls}
\end{eqnarray}
Substituting (\ref{cls}) into (\ref{drc}) and noticing
$S^{\mu\nu}_{ab}=-S^{\mu\nu}_{ba}$, we have
\begin{eqnarray}
\Re \lan\phi^+ S^{\mu\nu}_{a b}h^\al_{~n}\eta^{nb}\al^a\hat
p_\al\phi \ran\to m S^{\mu\nu}_{a b} u^a u^b \sqrt{1- v^2}\dl^3(\vec
x-\vec X)=0.\label{scls}
\end{eqnarray}
Omitting the tiny spin-gravity coupling energy, we get the usual EMT
for a classical particle with self-interactive potential
\begin{eqnarray}
T^{\mu\nu} \to( m u^\mu u^\nu+w g^{\mu\nu}- \Om_\al  s^\al
g^{\mu\nu})\sqrt{1- v^2}\dl^3(\vec x-\vec X).\label{clst}
\end{eqnarray}
$w>0$ acts like negative pressure\cite{gu1,gu3}. By (\ref{clst}) and
energy-momentum conservation law $T^{\mu\nu}_{~;\nu}=0$, we find
only if $w=0$ and  $m$ is a constant independent of $v$, the
particle moves along geodesic and the principle of equivalence
strictly holds\cite{gu2,ncs}.

Some previous works usually use one spinor to represent matter
field. This may be not the case, because spinor fields only has a
very tiny structure. Only to represent one particle by one spinor
field, the matter model can be comparable with general relativity,
classical mechanics and quantum mechanics\cite{gu1,gu2,gu3,gu4}. The
many body system should be better described by the Lagrangian
similar to the following one,
\begin{eqnarray}
 {\cal L}_m =\sum_n \Re  \lan \phi_n^+\wt\al^\mu \hat p_\mu\phi_n \ran +\Om_\mu \phi_n^+
\hat s^\mu \phi_n -m \phi_n^+\ga_0\phi_n+F_n,\quad F_n=\frac 1 2 \wt
w \ck\ga^2_n. \label{many}
\end{eqnarray}
The corresponding EMT is given by
\begin{eqnarray}
T^{\mu\nu} &=& \sum_n \l(\frac 1 2\Re \lan\phi_n^+(\wt\al^\mu \hat
p^\nu +\wt\al^\nu \hat p^\mu + 2 S^{\mu\nu}_{a
b}h^\al_{~c}\eta^{bc}\al^a\hat p_\al)\phi_n \ran  +
g^{\mu\nu}F_n\r. \nn\\
&~&\l. ~~~~- \frac 1 8 \eps^{abcd} h^\al_{~a} h^\be_{~b}
S^{\mu\nu}_{cd} (\pa_\al\ck s_{n\be}-\pa_\be \ck s_{n\al}) -
g^{\mu\nu}\Om_\al \ck s^\al_n\r). \label{drcm}
\end{eqnarray}
The classical approximation becomes
\begin{eqnarray}
T^{\mu\nu} \to\sum_n ( m_n u_n^\mu u_n^\nu+w_n g^{\mu\nu}- s_n^\al
\Om_\al g^{\mu\nu})\sqrt{1- v_n^2}\dl^3(\vec x-\vec
X_n),\label{clstm}
\end{eqnarray}
which leads to the EMT for average field of spinor fluid as follows
\begin{eqnarray}
T^{\mu\nu}= (\rho+P) U^\mu U^\nu +(W-P)  g^{\mu\nu}.\label{eos}
\end{eqnarray}
The self potential becomes the negative pressure $W$\cite{dm1}.

\section{Discussion and Conclusion}
\setcounter{equation}{0}

In this paper, according to the explicit relations between tetrad
and metric, we derived the exact representation of $T_{\mu\nu}$ of
spinors.  By splitting the spinor connection into geometrical part
$\Up_\mu$ and dynamical part $\Om_\mu$,  we find some new results.
In this EMT, besides the usual kinetic energy momentum term, the
nonlinear self-interactive potential $W$ acts like negative pressure
and may be important in astrophysics. The term $\Re\lan\phi^+
S^{\mu\nu}_{a b}h^\al_{~n}\eta^{nb}\al^a\hat p_\al \phi \ran$
reflects the interaction of momentum $p^\mu$ with tetrad, which
vanishes in classical approximation. The spin-gravity coupling term
is described by $\Om_\al s^\al$, which may be the origin for
magnetic field of a celestial body. This term also appears in the
dynamics of a spinor and has location and navigation functions for
the particle. To get a clear concept for vector $\Om_\al$, we make
weak field approximation for Kerr metric in cylindrical coordinate
system $(t,\rho,\vf,z)$. In this case, by (\ref{omOm}) we have
\begin{eqnarray}
ds^2 &\to&  dt^2- d\rho^2-\rho^2 d\vf^2- dz^2 - \frac {4ma \rho^2}
{r^3} dt d\vf,\qquad r=\sqrt{\rho^2+z^2}.\label{krm}\\
\Om^\al&=& \frac 1 {4\rho} (0,\pa_z g_{t\vf},0,-\pa_\rho g_{t\vf})=
\frac {ma}{2r^5} (0, 3\rho z,0, 2 z^2-\rho^2).
\end{eqnarray}
The vector $\Om^\al$ is similar to the magnetic field $\vec B$
generated by a circular coil. The lines of flux of vector $\Om^\al$
in spherical coordinate system are given by $r=R\sin^2\th,(R>0)$.
The spins of the particle will be arranged along these lines, and a
macroscopical magnetic field $\vec B$ can be induced. Since $dt$ is
not orthogonal to $d\vf$, this may lead to procession of the
magnetic field.

The decomposition of spin connection $\Ga_\mu$ not only makes
calculation simpler, but also highlights their different physical
meanings. Only in this representation, we can discover and clarify
the special effects as discussed above. In the calculation of tetrad
formalism we get a new tensor $S^{\mu\nu}_{ab}$ which plays an
important role in the interaction of spinor with gravity and appears
in many places. But it has not classical correspondences, and
vanishes in classical approximation.

The spinor has only a tiny but marvelous structure. It is an
indivisible system and unsuitable to describe many body problem.
Only by using one spinor to describe one particle, we get a harmonic
picture for field theory and classical mechanics.

\acknowledgments{ I'd like to acknowledge Prof. James M. Nester for
his enlightening discussions and encouragement. I have encountered
the difficulty in derivation of exact energy-momentum tensor for
many years. He encouraged ``There may be a deeper wisdom in these
things that we have not yet appreciated''. }


\begin{thebibliography}{99}
\bibitem{1} R. Rakhi, G. V. Vijayagovindan, and K. Indulekha, Int. Journ. Mod.
Phys. A 25, 2735 (2010).
\bibitem{2} M. O. Ribas, F. P. Devecchi, G. M. Kremer, {\em Fermions as sources of accelerated regimes in cosmology}, Phys. Rev. D72 (2005) 123502, ArXiv:gr-qc/0511099.
\bibitem{gu3} Y. Q. Gu, {\em A Cosmological Model with Dark Spinor Source},
Int. J. Mod. Phys. A22:4667-4678(2007), gr-qc/0610147
\bibitem{4} B. Saha, {\em Nonlinear Spinor field in isotropic space-time and dark energy models},  The European Phys. J Plus 131: 242 (2016)
\bibitem{5} R. Rakhi, G.V. Vijayagovindan, K. Indulekha, {\em A Cosmological Model with Fermionic Field}, Int. J. Mod. Phys. A25:2735-2746, 2010, arXiv:0912.1222
\bibitem{3} Ch. G. Boehmer, J. Burnett, D. F. Mota, D. J. Shaw,
{\em Dark spinor models in gravitation and cosmology}, JHEP
1007:053, 2010, arXiv:1003.3858
\bibitem{weyl} H. Weyl, {\em GRAVITATION AND THE ELECTRON}, PNAS  {\bf
15(4)} 323-334(1929)
\bibitem{Brill} D. R. Brill, J. A. Wheeler, Rev. Mod. Phys. {\bf 29(3)}, (1957)465
\bibitem{Weldon} H. Arthur Weldon, {\em Fermions without vierbeins in curved
space-time}, Phys. Rev. D63 (2001) 104010, arXiv:gr-qc/0009086
\bibitem{Zhang} H. B. Zhang, {\em Note on the EMT for general mixed tensor-spinor
fields},  Commun. Theor. Phys. 44 (2005) 1007-1010,
\bibitem{sachs} M. Sachs, {\em General relativity and matter}
{\bf (Ch.3)}, D. Reidel, 1982.
\bibitem{Newman} E. T. Newman, R. Penrose, J. Math. Phys. 3, 566(1962)
\bibitem{Bade} W. L. Bade, H. Jehle, Rev. Mod. Phys.{\bf  25(3)}, (1953)714
\bibitem{Bergm} P. G. Bergmann, Phys. Rev. {\bf 107(2)}, (1957)624
\bibitem{Crawf} J. P. Crawford, Adv. Appl. Cliff. Alg. {\bf 2(1)}, (1992)75
\bibitem{gu2} Y. Q. Gu, {\em The Simplification of Spinor Connection and Classical Approximation},
arXiv:gr-qc/0610001
\bibitem{nst}  J. M. Nester, Journal of Mathematical Physics 33, 910 (1992).
\bibitem{gu1} Y. Q. Gu, {\em Functions of State for Spinor Gas in General Relativity}, OALib. J., 4, e3953 (2017), arXiv:0711.1243
\bibitem{ncs} Y. Q. Gu, {\em Natural Coordinate System in Curved Space-time}, arXiv:gr-qc/0612176, 2017
\bibitem{gu4} Y. Q. Gu, {\em New Approach to N-body Relativistic Quantum Mechanics}, Int. J. Mod. Phys. A22:2007-2020(2007), hep-th/0610153
\bibitem{dm1} Y. Q. Gu, {\em Nonlinear Spinors as the Candidate of Dark
Matter}, OALib. J. 4, e3954 (2017), arXiv:0806.4649

\end{thebibliography}
\end{document}